\documentclass[preprint]{elsarticle}

\usepackage{graphicx}
\usepackage{amsmath}
\usepackage{color}
\usepackage{hyperref}
\usepackage{bigints}
\usepackage{relsize}
\newcommand{\ts}{\textstyle}
\newcommand{\gsim}{\raisebox{-0.6ex}{\mbox{ $\stackrel{\ts >}{\ts \sim}$ }}}
 \newcommand{\lsim}{\raisebox{-0.6ex}{\mbox{ $\stackrel{\ts <}{\ts \sim}$ }}}

\begin{document}
\title{Hydrodynamic model for picosecond propagation of laser-created nanoplasmas}
%\author{Vikrant Saxena$^{1,2}$, Zoltan Jurek$^{1,2}$, Beata Ziaja$^{1,2,3}$, and Robin Santra$^{1,2,4}$}
%\affiliation{
%$^1$ Center for Free-Electron Laser Science, DESY, Notkestrasse 85, 22607 Hamburg, Germany.\\
%$^2$ Hamburg Centre for Ultrafast Imaging, Luruper Chaussee 149, 22761 Hamburg, Germany.\\
%$^3$ Institute of Nuclear Physics, Polish Academy of Sciences, Radzikowskiego 152, 31-342 Krakow, Poland.\\
%$^4$ Department of Physics, University of Hamburg, Jungiusstrasse 9, 20335 Hamburg, Germany.}
\author[cfel,cui]{Vikrant Saxena}
\ead{vikrant.saxena@desy.de}
\author[cfel,cui]{Zoltan Jurek}
\author[cfel,cui,inp]{Beata Ziaja}
\ead{ziaja@mail.desy.de}
\author[cfel,cui,univHH]{Robin Santra}
\address[cfel]{Center for Free-Electron Laser Science, DESY, Notkestrasse 85, 22607 Hamburg, Germany}
\address[cui]{Hamburg Centre for Ultrafast Imaging, Luruper Chaussee 149, 22761 Hamburg, Germany}
\address[inp]{Institute of Nuclear Physics, Polish Academy of Sciences, Radzikowskiego 152, 31-342 Krakow, Poland}
\address[univHH]{Department of Physics, University of Hamburg, Jungiusstrasse 9, 20335 Hamburg, Germany}

\begin{abstract}
The interaction of a free-electron-laser pulse with a moderate or large size cluster is known to create a quasi-neutral nanoplasma, which then expands on hydrodynamic timescale, i.e., $>1$ ps. To have a better understanding of ion and electron data from experiments derived from laser-irradiated clusters, one needs to simulate cluster dynamics on such long timescales for which the molecular dynamics approach becomes inefficient. We therefore propose a two-step Molecular Dynamics-Hydrodynamic scheme. In the first step we use molecular dynamics code to follow the dynamics of an irradiated cluster until all the photo-excitation and corresponding relaxation processes are finished and a nanoplasma, consisting of ground-state ions and thermalized electrons, is formed. In the second step we perform long-timescale propagation of this nanoplasma with a computationally efficient hydrodynamic approach. 

In the present paper we examine the feasibility of a hydrodynamic two-fluid approach to follow the expansion of spherically symmetric nanoplasma, without accounting for the impact ionization and three-body recombination processes at this stage. We compare our results with the corresponding molecular dynamics simulations. We show that all relevant information about the nanoplasma propagation can be extracted from hydrodynamic simulations at a significantly lower computational cost when compared to a molecular dynamics approach. Finally, we comment on the accuracy and limitations of our present model and discuss possible future developments of the two-step strategy.
\end{abstract}

\maketitle 

\section{Introduction} 

The recently developed x-ray free-electron lasers (FEL) \cite{FLASH,LCLS,SACLA} open up new horizons in the experimental investigation of laser-matter interaction. In particular, the study of FEL irradiated atomic clusters   \cite{Thomas2013PRL,Gorkhover2012PRL} is an important step towards the understanding of  the behavior of radiation-induced ionization dynamics within more complex systems. The FEL induced dynamics depend on the cluster size. After FEL irradiation, a small cluster transforms into a positively charged ion cloud. In contrast, a large irradiated cluster transforms into a quasi-neutral  nanoplasma \cite{hau-riege}. In both cases, the system finally expands, however, the expansion dynamics are different. In the first case, the positively charged ion cloud explodes due to the unscreened electrostatic repulsion between ions, whereas in the second case the quasi-neutral nanoplasma expands due to the thermal pressure of electrons, which occurs on a much slower timescale. The expansion of quasi-neutral plasma into vacuum has been a subject of continuous research interest \cite{mora1,mora2,mora3,murakami,peano,beck,popov,kiefer}. Among them, Mora has discussed the cases of isothermal \cite{mora1,mora2}, as  well as adiabatic \cite{mora2,mora3} expansion, while assuming a Maxwellian distribution for electrons with a homogeneous temperature. Murakami et al. \cite{murakami} have used a two-fluid model (later also a three-fluid model with two electron fluids of different temperatures) without the assumption of Maxwellian electron distribution. They have obtained analytical self-similar expressions for the ion front position and maximum ion energy by assuming a polytropic relation between electron temperature and density. In more recent works, authors have either used a Vlasov simulation \cite{peano,popov} or an N-body simulation \cite{beck} approach. Popov et al. \cite{popov} have also  investigated the case of different initial density profiles. Kiefer et al. \cite{kiefer} have shown that the use of a hydrodynamic model without the assumption of a Maxwellian electron energy distribution yields more accurate  estimates for the ion front position and for the maximum ion kinetic energy than its Maxwellian counterpart.

On the other hand, in order to understand spectroscopic data (ion charge states, ion kinetic energies, electron kinetic energies) from FEL-induced cluster dynamics, long-timescale ($\gsim 1$ ps) computer simulations are required. With particle approaches, this is computationally expensive.  The existing literature on classical molecular dynamics (MD) simulations of FEL irradiated clusters is therefore limited to small clusters ($N\lsim10^4$) \cite{jurek1,C60LCLS,ArClusterSACLA}. Only in a recent study by Peltz et al. in Ref. \cite{peltz2014} was the NIR-XFEL irradiation of a large cluster consisting of $\sim 10^6$ atoms investigated; but, these studies were limited to relatively short timescales $\sim 100$ fs. The MD simulation of the atomic cluster with $10^6$ atoms was carried out on a $80$ CPU cluster. The run took on the order of $30$ days \cite{peltz2014}.

We therefore propose a computationally efficient two-step MD-Hydrodynamic approach for simulating the picosecond relaxation of irradiated large samples. The first step consists of MD simulations performed during and after the XFEL pulse until all photo-excitation and the corresponding relaxation processes are finished, i.e., until all ions and atoms are in ground state configurations and electrons are in local thermal equilibrium. The second step consists of propagating the ground state ions and the thermalized electrons on a picosecond timescale with a hydrodynamic code. At the present stage of development of our hydrodynamic code only collisionless propagation of electron and ion densities are employed to test the accuracy and stability of our code. In the next development step we plan to include collisional processes (electron impact ionization and three-body recombination) into our approach.

For our simulations we consider a quasi-neutral plasma consisting of singly ionized argon ions ($\sim 10^5$ atoms) and the same number of plasma electrons. A two-fluid approach (involving a warm electron fluid intermixed with a cold ion fluid) is used. Our approach is different from that of Hau-Riege et al. \cite{hau-riege} as they solve the integrated rate equations for electron and ion densities whereas here we solve full partial differential equations for local densities at each grid point. Our model is closer to that of Murakami and Basko \cite{murakami}, except that they use a polytropic relation between temperature and density. Our hydrodynamic simulations are compared with the corresponding molecular dynamics simulations which were performed using the GPU accelerated version of our in-house developed molecular dynamics code, XMDYN  \cite{xmdyn}. We show that the hydrodynamic approach can accurately reproduce the propagation of a nanoplasma at a much lower computational cost than molecular dynamics. 

In the next section, we present our one-dimensional two-fluid model for the propagation of a spherically-symmetric net-neutral nanoplasma. In Sec.~3 the methodology of the hydrodynamic simulations and the molecular dynamics simulations is described briefly. In Sec.~4 we present our hydrodynamic simulation results for an argon nanoplasma and discuss the effect of initial electron temperature on the nanoplasma propagation. In Sec.~5 a comparison of our hydrodynamic simulation results with the results of our molecular dynamics simulations is made. Limitations of both approaches are discussed. Finally in Sec.~6 we present a summary. 

%%%%%%%%%%%%%%%%%%%%%%%%%%%%
\section{Two-Fluid Model for a Spherically Symmetric Nanoplasma}

We consider a spherically symmetric nanoplasma and assume that the symmetry of the sample is preserved at all times. The two-fluid model equations for a system of a cold ion fluid intermixed with a warm electron fluid can be derived from the kinetic plasma equations by a straightforward integration over the velocity space. While assuming spherical symmetry, the final equations take the form,

\begin{equation}
\frac{\partial (r^2n_e)}{\partial t}+\frac{\partial\left(r^2n_eu_e\right)}{\partial r} = 0
\label{el_cont_1}
\end{equation}  
\begin{equation}
\frac{\partial (r^2n_e u_e)}{\partial t}+ \frac{\partial\left(r^2 n_e u_e^2 \right)}{\partial r} = 
\frac{e(r^2n_e)}{m_e}\frac{\partial \phi}{\partial r} - \frac{r^2}{m_e}\frac{\partial \left(n_e T_e\right)}{\partial r} 
\label{el_moment_1}
\end{equation}  
\begin{equation}
\left(\frac{\partial}{\partial t}+u_e\frac{\partial}{\partial r}\right)T_e = -\frac{2}{3}T_e 
\left(\frac{1}{r^2}\frac{\partial \left(r^2 u_e\right)}{\partial r}\right)
\label{Te_pde}
\end{equation}  
\begin{equation}
\frac{\partial (r^2n_i)}{\partial t}+\frac{\partial\left(r^2n_iu_i\right)}{\partial r} = 0
\label{ion_cont_1}
\end{equation}  
\begin{equation}
\frac{\partial (r^2n_i u_i)}{\partial t}+ \frac{\partial\left(r^2 n_i u_i^2\right)}{\partial r} = 
-\frac{Ze(r^2n_i)}{m_i}\frac{\partial \phi}{\partial r}
\label{ion_moment_1}
\end{equation}  
\begin{equation}
\frac{1}{r^2}\frac{\partial}{\partial r} \left(r^2 \frac{\partial\phi}{\partial r}\right)= \frac{e(n_e-Z\,n_i)}{\epsilon_0}
\label{pois}
\end{equation}

Here symbols $n$, $u$, and $T$ stand for the fluid density, fluid velocity and fluid temperature respectively whereas the subscript $e$ ($i$) represents the electron (ion) fluid. The electrostatic potential is denoted by the symbol $\phi$, $e$ stands for the magnitude of the electronic charge, $Z$ is the charge state of the ion fluid and $\epsilon_0$ is the vacuum permittivity, $m_e$ and $m_i$ are the masses of electron and ion respectively. The radial position is represented by $r$ and $t$ denotes time. The equations are defined as: Eq.~(\ref{el_cont_1}), is the continuity equation for the electron fluid and describes the conservation of electron number; Eq.~(\ref{el_moment_1}) describes the conservation of the electron fluid momentum; Eq.~(\ref{Te_pde}) governs the time evolution of the electron fluid  temperature. The negative sign on the right hand side indicates that the temperature of the electron fluid tends to decrease as the fluid expands into the vacuum; the equations Eq. (\ref{ion_cont_1}) and Eq. (\ref{ion_moment_1}) are continuity and momentum equations for the ion fluid, respectively.  The absence of the pressure term in the continuity equation for the ion fluid is due to the assumption that the ion fluid remains cold at all times ($T_i=0$ eV). The electron fluid on the other hand has a finite time-dependent temperature ($T_e>0$ eV) unlike the case of isothermal expansion \cite{mora1}, wherein the electron fluid is assumed to instantly achieve thermodynamic equilibrium and to retain the initial temperature. Here we account for electron escape and therefore the electron fluid remains in local thermodynamic equilibrium only. Hence, we allow the total electron temperature to evolve during the expansion, but we assume that the temperature across the simulation box remains uniform. This simplifying assumption is consistent with previous hydrodynamic models of nanoplasma expansion. With this, the equation for the temperature evolution can be reduced to the following form:

\begin{equation}
\frac{\partial T_e}{\partial t} = -\frac{2}{3}T_e 
\frac{\mathop{\mathlarger{\bigintssss n_e\frac{\partial}{\partial r} \left(r^2 u_e\right) dr}}}
{\mathop{\mathlarger{\bigintssss r^2 n_e dr}}}
\label{Te_ode}
\end{equation}  

The spatial profile of the electrostatic potential depends on the charge density distribution according to Poisson's equation, Eq.(\ref{pois}). The space-charge field which can be readily obtained from the electrostatic potential governs the dynamics of the charged fluids along with the electron thermal pressure (second term on the right hand side of the electron momentum equation, Eq. (\ref{el_moment_1})). In order to follow nanoplasma expansion, we need to solve the coupled set of  time-dependent partial differential equations, i.e.,  Eqs.  (\ref{el_cont_1},\ref{el_moment_1},\ref{ion_cont_1},\ref{ion_moment_1}) along with Eq. (\ref{pois}) and Eq. (\ref{Te_ode}).  In the next section we discuss the details of the numerical methods used for our simulations.

%%%%%%%%%%%%%%%
\section{Numerical Methods}

\subsection{Hydrodynamics: \textit{XHYDRO}}

The ordinary differential equation for the time evolution of electron temperature is solved using a fourth-order Runge-Kutta scheme \cite{num_recipe}. The other time-dependent partial differential equations are in the form of continuity equations and are solved using the LCPFCT subroutine package \cite{lcpfct} which is based on the flux-corrected-transport (FCT) scheme \cite{boris}. The Poisson equation is discretized using a finite difference method and is solved using a standard tridiagonal solver \cite{num_recipe}. For achieving stability as well as efficiency at the same time we have used an adaptive time step that satisfies the Courant stability criterion throughout the simulation.

It is well known that the occurrence of shock fronts is inevitable in expansion dynamics as investigated here. In general, these shock fronts are difficult to resolve but there are some existing recipes how to deal with them, e.g., through the application of artificial viscosity. The concept of artificial viscosity was first introduced by Richtmyer \cite{richtmyer,neumann} and has been further investigated in several other works \cite{laxwendroff,lapidus,lohner,nessyahu,bianchini}. Here we follow the formulation of Lapidus \cite{lapidus,lohner} and use the artificial viscosity term in all the conservation law equations. This modifies the equations, Eqs.(\ref{el_cont_1})-(\ref{ion_moment_1}) as follows.

\begin{equation}
\frac{\partial (r^2n_e)}{\partial t}+\frac{\partial\left(r^2n_eu_e\right)}{\partial r} = 
\frac{\partial}{\partial r} F\left(u_e,r^2n_e\right)
\label{el_cont_2}
\end{equation}  
\begin{eqnarray}
\frac{\partial (r^2n_e u_e)}{\partial t}+ \frac{\partial\left(r^2 n_e u_e^2\right)}{\partial r} 
&=& \frac{e(r^2n_e)}{m_e}\frac{\partial \phi}{\partial r} - \frac{r^2}{m_e}\frac{\partial \left(n_e T_e\right)}{\partial r} 
\nonumber \\
&+& \frac{\partial}{\partial r} F \left(u_e,r^2n_eu_e\right)
\label{el_moment_2}
\end{eqnarray}  
\begin{equation}
\frac{\partial (r^2n_i)}{\partial t}+\frac{\partial\left(r^2n_iu_i\right)}{\partial r} = 
\frac{\partial}{\partial r} F\left(u_i,r^2n_i\right)
\label{ion_cont_2}
\end{equation}  
\begin{eqnarray}
\frac{\partial (r^2n_i u_i)}{\partial t}+ \frac{\partial\left(r^2 n_i u_i^2\right)}{\partial r} 
&=& -\frac{Ze(r^2n_i)}{m_i}\frac{\partial \phi}{\partial r}\nonumber \\  
&+& \frac{\partial}{\partial r} F\left(u_i,r^2n_iu_i\right)
\label{ion_moment_2}
\end{eqnarray}  

In all the above equations, the right hand side term $F(A,B)=c_1(dr)^2\left|\frac{\partial A}{\partial r} \right| \frac{\partial B}{\partial r}$, where $dr$  stands for the numerical grid spacing, is the artificial viscosity term and can be controlled by choosing an appropriate value of the coefficient $c_1$. It is to be emphasized that the use of artificial viscosity causes a loss of total energy during the propagation and also affects the structure of the sharp wave fronts. Therefore one needs to choose an artificial viscosity that is as low as possible. This requires a series of numerical tests. As a result, we have used a value $c_1=1$ for all simulations presented here. 

With the viscosity terms included, the above numerical scheme is complete. In what follows we will refer to it as to the XHYDRO code.

%%%%%%%%%%%%%%%%%%%%%%%%%%%%%%%%%%%%%%%%%%%%%%
\subsection{Molecular dynamics code for validating hydrodynamic simulations}

XMDYN \cite{xmdyn} is a versatile molecular dynamics simulation code for modeling matter irradiated by intense x-ray pulses. The approach has been validated against various experimental results \cite{C60LCLS,ArClusterSACLA,JPHYSB2014}. The code follows the ionization dynamics of the individual atoms initiated by x-ray photons using the Monte Carlo method. The required atomic parameters are calculated by the XATOM toolkit \cite{xatom}. The real space propagation is done by describing the atoms, ions and ionized electrons as classical point-like particles interacting via classical forces, e.g., Coulomb forces between charges. The velocity Verlet algorithm is applied to numerically integrate Newton's equations for particles. A regularized form of the Coulomb potential $~1/ \sqrt{(r^2+r_0^2)}$, with a cutoff parameter $r_0$ of a fixed value, is used to remove the  Coulomb singularity and ensure numerical stability. The long-range Coulomb forces are calculated for all particle-particle pairs explicitly. In order to accelerate this time-consuming force evaluation, a GPU extension based on Ref. \cite{CUNBODY} was used. In the current work the photoinduced and collisional ionization processes within XMDYN were switched off and only the propagation scheme was left on as the collisionless propagation of a electron-ion plasma was in focus of our study.

%%%%%%%%%%%%%%%%%%%%%%%%%%%%%%%%%%%%%%%%%%%%%%%
\section{Hydrodynamics simulations of nanoplasma expansion and their comparison with MD results}

For the hydrodynamic simulations presented here we chose an initial condition corresponding to a quasi-neutral spherical nanoplasma of radius $100$ \AA$\,$ which consists of a positively charged cold fluid ($T_i=0$ eV) of singly charged Ar ions intermixed with a warm electron fluid ($T_e>0$ eV). The density of the argon cluster is considered to have a realistic value of $1.9742\times10^{28}$ m$^{-3}$ corresponding to an interatomic separation of $3.7$ \AA. For all simulation runs we have considered a simulation box size of $R_{max} = 1000$ \AA$\,$ with the grid resolution fixed to  $dr = 0.1$ \AA. The initial time step is taken as $dt = 0.01$ attoseconds, whereas adaptive time steps are used at later times. We use reflective boundary condition on the left boundary (symmetry axis at $r=0$), whereas an open boundary condition is used at the right boundary ($r=R_{max}$). The set of conservation law equations, Eqs.~(\ref{el_cont_2})-(\ref{ion_moment_2}) is then numerically integrated along with Eq.~(\ref{pois}) and Eq.~(\ref{Te_ode}). The simulation results are discussed below.

In Fig.1 we present the XHYDRO results for the case of $T_{e0}= 33$ eV ($T_{e0}\equiv T_e(t=0)$). They follow the expansion of electron and ion clouds. Plots of radial densities of electron and ion fluids are shown at $t=0$ ps,  $t=1$ ps and $t=2$ ps. It should be noted that in the centre of the spherically symmetric cloud the plasma remains neutral and the space charge field is localized around the ion front. It indicates that the hydrodynamic expansion is the dominant expansion mechanism for our system. Moreover, it is also observed that in the outer region the electron radial density, $r^2n_e$, vanishes with increasing $r$, whereas the ion density front develops a knee-like structure which rushes outwards as time progresses. However, with increasing time the charge separation and thus the space charge field at the front diminishes. In Fig.~2 the radial positions of ion fronts at $t=1$ ps for three different values of initial electron temperature, $T_{e0}=33$ eV, $60$ eV and $100$ eV are shown. The initial configuration and therefore the initial ion front position is the same for the three different temperature runs. It is observed that the ion front position strongly depends on the initial electron temperature via the ion acoustic velocity whereas the ion density at the knee is weakly dependent on the initial electron temperature.

\begin{figure}[!ht]
\includegraphics[scale=0.5]{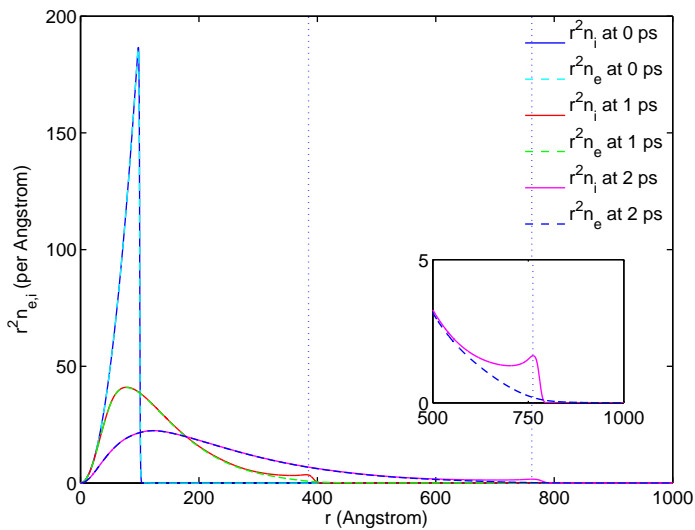}
\caption{(color online) Expansion of electron and ion fluids for $T_{e0}=33$ eV on a picosecond time scale.}
\end{figure}

\begin{figure}[!ht]
\includegraphics[scale=0.5]{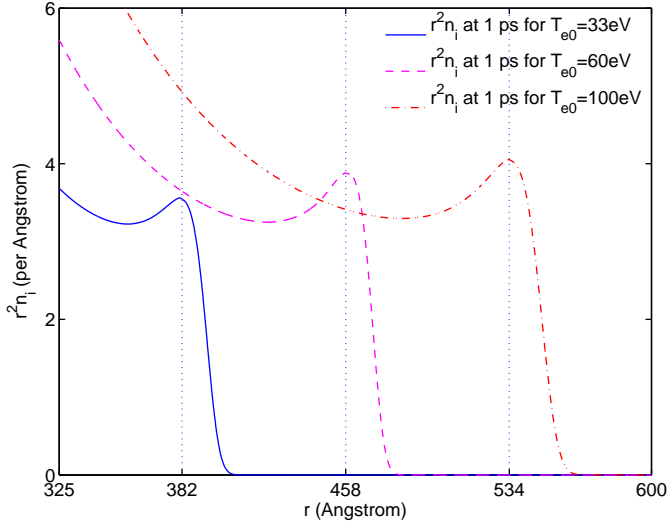}
\caption{(color online) Effect of initial electron fluid temperature, $T_{e0}$, on the position and on the  magnitude of the ion front after $1$ ps expansion.}
\end{figure}

For the molecular dynamics simulations, a single charged $100$ \AA-large argon cluster of density $1.9742\times10^{28}$ m$^{-3}$ with a net-neutralizing electron cloud corresponds to $82712$ singly charged Ar ions with the same number of plasma electrons. The electrons first thermalize with each other (in the presence of cold single charged ions) to reach the required initial electron temperature corresponding to that of the hydrodynamic simulation. The dynamics of these $2N=165424$ particles is then followed by our MD tool. The N-particle simulation on the ps time scale is quite challenging due to the $N^2$ scaling of the required simulation time with MD. We comment on this aspect in Sec.~5.2. 

In what follows we compare the hydrodynamic simulation results with the molecular dynamics simulation performed on a GPU (one realization). In Fig. 3 and Fig. 4 a comparison of the electron and ion radial density profiles at $t=1$ ps for $T_{e0}=33$ eV and $60$ eV is shown. 

\begin{figure}[!ht]
\includegraphics[scale=0.5]{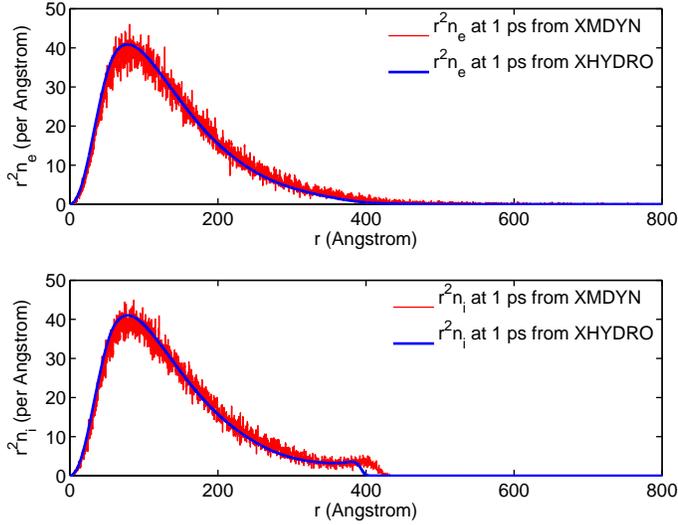}
\caption{(color online) Comparison of  radial densities of electrons (upper panel) and ions (lower panel) at $t=1$ ps  obtained from XMDYN and XHYDRO simulations for the initial electron temperature of $T_{e0}=33$ eV.}
\end{figure}

\begin{figure}[!ht]
\includegraphics[scale=0.5]{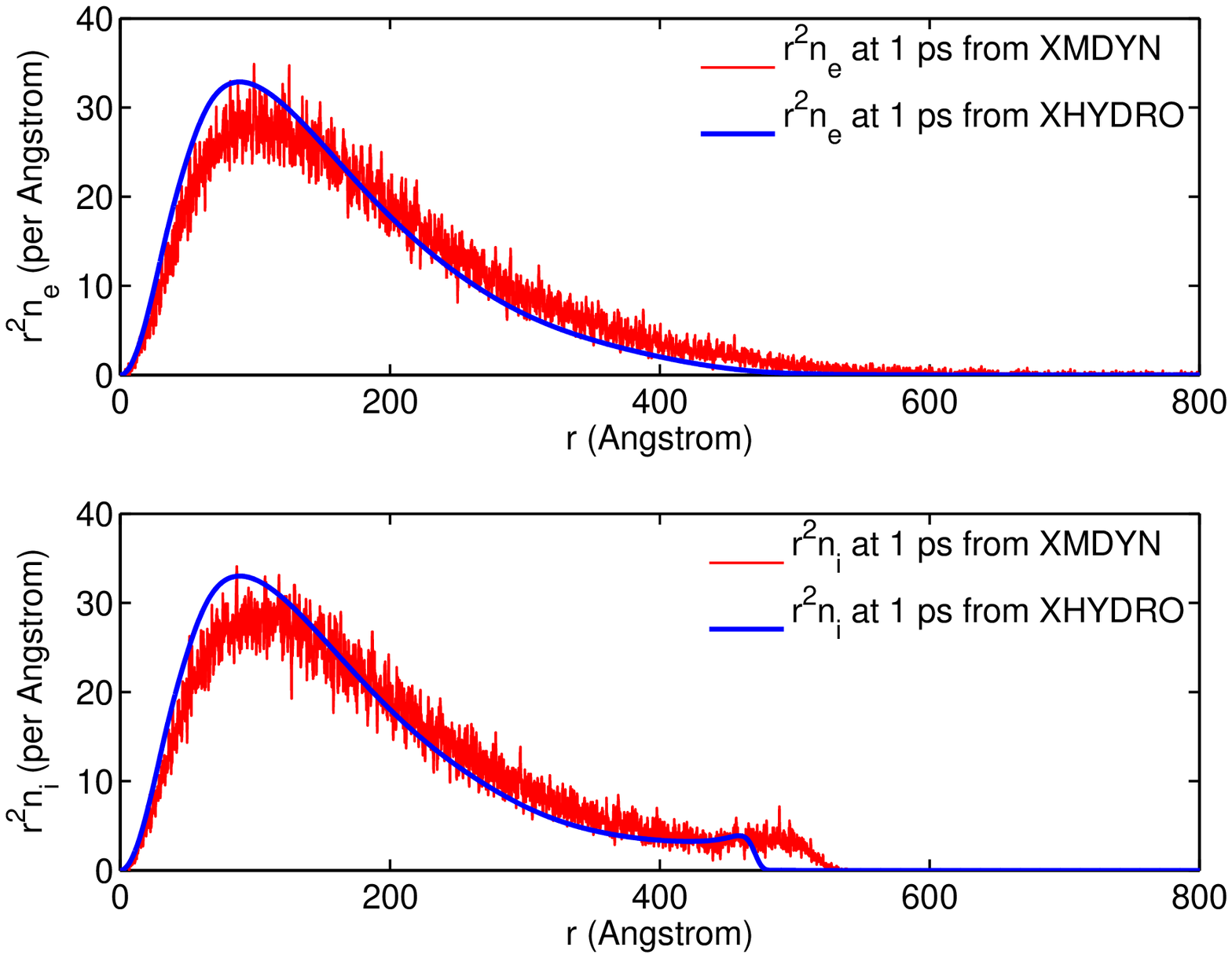}
\caption{(color online) Comparison of radial densities of electrons (upper panel) and ions (lower panel) at $t=1$ ps  obtained from XMDYN and XHYDRO simulations for the initial electron temperature of $T_{e0}=60$ eV.}
\end{figure}

The densities obtained with the two different simulation techniques are in a good agreement except that they predict a slightly different position for the ion front. To understand this discrepancy we also investigated the time evolution of electron temperature within the simulation box. The time evolution of electron temperature calculated with the two approaches is shown in Fig. 5. It is found that the hydrodynamic approach predicts a faster decrease of the electron temperature during the early phase of expansion. This decrease is due to the loss of initially emitted very hot photoelectrons from the simulation box in the absence of the electron heat flux (due to the assumption of uniform temperature within our model). Eventually, the resulting lower temperature of electron fluid yields a lower thermal pressure in the hydrodynamic case than in the molecular dynamics case. Due to the lower thermal pressure the plasma expansion becomes slower. Therefore we can see that in the hydrodynamic case the electron front and the ion front lie behind their counterparts in the molecular dynamics case. 

\begin{figure}[!ht]
\includegraphics[scale=0.5]{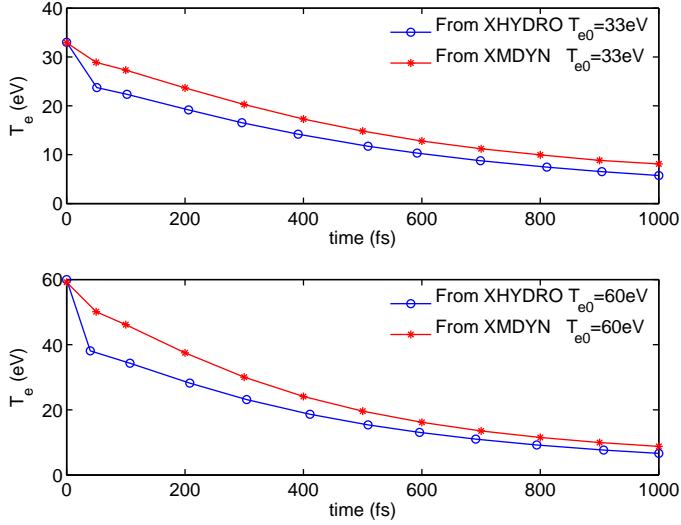}
\caption{(color online) Evolution of electron temperature obtained with XHYDRO and XMDYN codes for the initial electron temperatures of $T_{e0}=33eV$ (upper plot) and $T_{e0}=60eV$ (lower plot).}
\end{figure}

%%%%%%%%%%%%%%%%%%%%%%%%%%%%%%%%%%%%%%%%%%
\section{Accuracy and Efficiency of XHYDRO as compared to XMDYN}

\subsection{Conservation laws: accuracy vs stability}

For solving the set of hydrodynamic equations, Eqs.(\ref{el_cont_2})-(\ref{ion_moment_2}), we have used a subroutine package \cite{lcpfct} based on flux-corrected transport algorithm which involves a diffusion and anti-diffusion stage. By choosing anti-diffusion to be slightly lower than the diffusion an extra diffusion coefficient can be added to ensure smooth field profiles. After numerical tests we have chosen a residual diffusion coefficient of $0.995$ (a value of $1.0$ corresponds to no additional diffusion). This introduces a dissipation of the energy out of the system. Moreover, we have used an artificial viscosity in our fluid equations, Eqs.(\ref{el_cont_2})-(\ref{ion_moment_2}) to stabilize the discontinuities (shocks), which also contributes to the loss of energy. In addition, energy is carried away by the fast electrons that leave the box in the early stage of the expansion. Therefore, in our hydrodynamical simulations we expect a reduction of the electron number and of the total energy of the system, especially in the initial stage of the sample evolution. The expressions for total energy in the MD and hydrodynamics schemes are given below. In MD, the total energy can be written as:

\begin{equation}
E_{total}^{MD} = \sum_{i=1}^{2N}{\frac{1}{2}m_{i} v_{i}^2}+
\frac{1}{2}\sum_{i=1}^{2N}{\sum_{j\neq i,j=1}^{2N}{\frac{1}{4\pi\epsilon_0} \frac{q_iq_j}{r_{ij}^2}}}  
\end{equation}

On the other hand, in hydrodynamics, thermal motion is separated from the fluid motion. Therefore the total energy reads:

\begin{equation}
E_{total}^{Hydro} = \int  {4\pi r^2\left(\frac{1}{2}\left(n_em_eu_e^2+n_im_iu_i^2\right)+\frac{3}{2}n_eT_e+\frac{1}{2}\epsilon_0\left(\frac{\partial \phi}{\partial r}\right)^2\right)dr    }
\end{equation}

In Fig. 6 we present particle number conservation and the total energy conservation within a $1000$ \AA-large simulation box for hydrodynamic simulation as compared with MD simulation. We see a larger loss of electrons with XHYDRO as compared to XMDYN ($\sim 1\%$ relative difference between final values) that is due to the use of artificial viscosity and numerical diffusion while solving the continuity equation. Moreover, as the fastest electrons leave the simulation box due to their thermal motion, they carry some energy away. This energy loss together with the dissipation losses due to numerical diffusion and artificial viscosity account for significantly larger energy loss in XHYDRO as compared to the energy loss observed in XMDYN ($\sim 8\%$ relative difference between the final values). It should also be mentioned here that the initial state of the XHYDRO simulation is quasi-neutral. In XMDYN there is a non-zero electrostatic energy present at the beginning of the simulation due to the particle granularity.

\begin{figure}[!ht]
\includegraphics[scale=0.5]{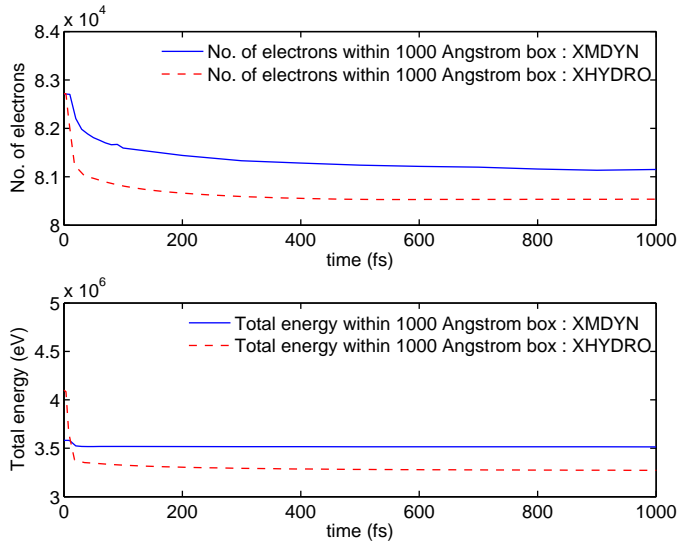}
\caption{(color online) Comparison of the particle number conservation and total energy conservation within a $1000$ \AA$\,$  box, in XHYDRO and in XMDYN for the case of $T_{e0} = 33eV$.} 
\end{figure}

\subsection{Efficiency}

Although particle methods are more accurate, they become computationally inefficient for large systems. This is because the computational cost in MD, when computing all pairwise interactions (as is currently done in XMDYN), scales as $N^2$, $N$ being the number of particles to be simulated.  We have estimated that the CPU version of XMDYN would need more than a year to simulate the propagation of $2*82712 = 165424$ particles during a picosecond. Using the GPU acceleration the computational time can be reduced down to approximately 3 days on a single GPU. This is still longer than the computational requirement for an equivalent hydrodynamics simulation which needs $< 10$ hours on a single CPU. Further estimates show that even if the forces in MD were evaluated with a tree method (which scales as $N\, *\, log\,N$), the hydrodynamic approach to follow the expansion of nanoplasmas involving a large number of particles, $N \gsim 10^5$, would be still more efficient than MD. It should be emphasized here that the above computational time estimates are for only one MD realization of the sample evolution. For realistic cases, runs of different realizations are performed which are then ensemble averaged. In hydrodynamics we do not need this procedure, as ensemble averaging is inherent in the construction of the hydrodynamics equations. This makes the hydrodynamic approach even more computationally efficient. Generally, the computational cost of the hydrodynamic simulation mostly depends on the number of grid points used. This implies that systems of various sizes can be investigated with the same computational efficiency as long as they 'fit' within the one simulation box of a fixed size. If the simulation box size has to be modified, the computational time for hydrodynamics simulation increases approximately linearly with the box size. In addition, the adaptive time step used in hydrodynamic simulations scales inversely with the maximum velocity present in the system and therefore it decreases with the initial electron temperature as $\propto 1/\sqrt{T_{e0}}$. Nevertheless, in the case considered here the overall computational costs of hydrodynamic scheme for large systems remain much lower that those of the MD-based models.

Let us emphasize that the imposed condition of spherical symmetry for the nanoplasma evolution in the hydrodynamics approach outlined here reduced the calculations to 1D case.  This assumption holds, e.g., for nanoplasma created from atomic clusters. In general, for an asymmetric sample a full three-dimensional hydrodynamics simulation would be required. The 3D calculation would be more expensive computationally than its spherically symmetric counterpart, due to the scaling of computational efficiency with the number of grid points. A dedicated study of efficiency of the 3D hydrodynamic approach as compared to the corresponding molecular dynamics simulations should then be performed.

%%%%%%%%%%%%%%%%%%%
\section{Summary}

We have presented a numerical study of the expansion of a net-neutral spherically symmetric nanoplasma (consisting of Ar$^{+1}$ ions and plasma electrons) based on a 1D hydrodynamic model. The argon ions were assumed to remain cold throughout the simulations whereas the electron temperature was allowed to evolve in time. The hydrodynamic results have been compared with molecular dynamics simulations carried out on GPU architecture. We have shown that the hydrodynamic calculations closely reproduce the nanoplasma propagation predicted by MD simulations; but, are performed at much lower computational costs than are the MD calculations. 

For realistic hydrodynamic simulations of the nanoplasma dynamics on the picosecond timescale, the inclusion of rate equations for three-body recombination and impact ionization is still required. After performing this, the computationally efficient two-step MD-Hydrodynamic approach for simulating picosecond evolution of irradiated large samples would be complete and could be applied for analysis of experimental data. This effort is planned and will be reported elsewhere.

%%%%%%%%%%%%%%%
\section{Acknowledgements}

VS thanks Markus Becker of INP Greifswald, Germany for his useful suggestions on hydrodynamic shock simulations. Authors are grateful to R. W. Lee for helpful comments.
%--------------------------------------------

%%%%%%%%%%%%%%%%%%%%%%%%%%%%%%%%%%%%%%%%%%%%%%%%%%%%%%%%%%%%%%%%%%%%%
\end{document}